\documentclass[twocolumn]{aastex631}
\makeatletter
\renewcommand{\frontmatter@title@above}{}
\makeatother
\usepackage{amsmath}
\usepackage{amssymb}
\usepackage{bm}
\usepackage{booktabs}
\usepackage{array}
\usepackage{graphicx}
\usepackage{microtype}
\usepackage{makecell}
\usepackage{appendix}

\shorttitle{Is Dark Energy Dynamical? A Symbolic Regression Analysis}
\shortauthors{Mouta et al.}

\begin{document}

\title{Is Dark Energy Dynamical? A Symbolic Regression Analysis}

\author{Andressa Mouta}
\affiliation{Observatório Nacional, Rua General José Cristino, 77, 20921-400, Rio de Janeiro - RJ, Brazil}
\author{Agripino Sousa-Neto}
\affiliation{Observatório Nacional, Rua General José Cristino, 77, 20921-400, Rio de Janeiro - RJ, Brazil}

\author{Jailson Alcaniz}
\affiliation{Observatório Nacional, Rua General José Cristino, 77, 20921-400, Rio de Janeiro - RJ, Brazil}


\begin{abstract}
Recent cosmological observations have provided intriguing indications of dynamical dark energy, motivating approaches that can explore its evolution without imposing specific functional forms. In this work, we investigate the redshift dependence of the normalized dark-energy density, $X(z)\equiv\rho_{\rm DE}(z)/\rho_{\rm DE,0}$, using Exhaustive Symbolic Regression (ESR). Our approach systematically explores analytical expressions within a prescribed space of mathematical operators and symbolic complexity, subject to physically motivated constraints. We constrain the resulting expressions using Type Ia supernovae, baryon acoustic oscillation, and cosmic microwave background observations, and assess their statistical performance and complexity through the Minimum Description Length (MDL) principle. Within the explored symbolic space and the adopted MDL criterion,
the constant-density solution $X(z)=1$ is preferred, while several
nonconstant expressions remain observationally viable, highlighting
the limited ability of current cosmological observations to
constrain the evolution of dark energy. 
\end{abstract}

\keywords{cosmology, dark energy, observations, data analysis, symbolic regression}


\section{Introduction}
\label{sec:introduction}

Recent baryon acoustic oscillation (BAO) measurements from the Dark Energy
Spectroscopic Instrument (DESI), combined with cosmic microwave background
(CMB) and Type Ia supernova (SN Ia) observations, have renewed interest in
the possibility that dark energy evolves with cosmic time. In particular,
the DESI Data Release 2 analysis found a preference for a time-dependent
dark-energy equation of state over a cosmological constant in several
data combinations \citep{DESI2025}. Updated SN Ia analyses, including
DES-Dovekie and Union3.1, have reduced the reported significance from as
high as $4.2\sigma$ to approximately $3.2$--$3.4\sigma$, illustrating both
the persistence of the indication and its sensitivity to the treatment of
observational data \citep{Popovic2026,Hoyt2026,LahavShah2026}. Moreover, recent analyses have shown that physically motivated constraints on the age of the Universe can substantially weaken the preference for dynamical dark energy, bringing the inferred evolution into closer agreement with a cosmological constant \citep{Sousa-Neto:2026zsb}. If confirmed,
dynamical dark energy would challenge a central assumption of the standard
Lambda Cold Dark Matter ($\Lambda$CDM) model and could point to new physics in the dark sector.
Determining whether the observed preference reflects genuine cosmic evolution, rather than modeling assumptions, choices of observational data, or residual systematic effects, therefore remains an important open problem.

Most of these indications arise from phenomenological descriptions of the
dark-energy equation of state $w(z)$
\citep{Chevallier2001,Linder2003,Barboza2008}. Moreover, some of these phenomenological analyses favor equation of state histories that cross the phantom divide, $w =-1$. Such behavior cannot be realized by a minimally coupled quintessence scalar field \citep{LahavShah2026}. At the same time, non-parametric and
model-independent studies have not established robust departures from
$\Lambda$CDM \citep{Dinda2025,Sousa-Neto:2025gpj}. Taken together, these results
motivate methods capable of exploring a wider range of expansion histories
without adopting a specific dark-energy equation of state from the outset.

The choice of the reconstructed quantity is also important. The effect of
$w(z)$ on cosmological distances is mediated by the integral relating the
equation of state to the dark-energy density and by the further integration
of the inverse expansion rate. These successive integrations smooth
features in $w(z)$ and make distinct histories difficult to discriminate
observationally. The normalized dark-energy density,
$X(z)$ instead enters directly into 
the Friedmann equation, offering a more immediate description of its
contribution to the background expansion. This is especially relevant
because some phenomenological fits imply non-monotonic density evolution,
including a rise followed by a decline \citep{LahavShah2026}. Reconstructing
$X(z)$ allows such behavior to be tested rather than imposed. Recent
studies of the dark-energy density have also found that current
data remain compatible with a cosmological constant \citep{deSouza2026,WangFreese2026}.

Symbolic Regression (SR) provides a complementary way to investigate this question by treating the mathematical expression itself as an object of inference. SR is a data-driven approach that seeks to identify analytical expressions describing the relationships between variables without assuming a predefined functional form. Unlike conventional regression methods, SR simultaneously searches for the mathematical structure and numerical parameters of the underlying function. Several approaches have been developed, including Eureqa \citep{SchmidtLipson2009}, AI Feynman \citep{UdrescuTegmark2020}, and PySR \citep{cranmer2023interpretable}, with applications extending to cosmology \citep{Arjona_2020}. In this work, we employ Exhaustive Symbolic Regression (ESR), which systematically enumerates expressions within a specified operator basis and complexity limit \citep{Bartlett_ESR}. The resulting candidates can be evaluated both by their agreement with observations and by their mathematical simplicity. This balance is particularly relevant to the current dark-energy debate: any proposed evolution must justify the extra descriptive complexity relative to the constant-density prediction of $\Lambda$CDM \citep{LahavShah2026}.

In this work, we develop an ESR framework to reconstruct the redshift
evolution of the normalized dark-energy density and use Type Ia supernova and BAO observations, supplemented by a compressed Gaussian prior derived from CMB information to constrain our resulting expressions.
We combine exhaustive enumeration with physically motivated admissibility conditions,
parameter estimation, and a 
Minimum Description Length (MDL) criterion \citep{Grunwald2007} that
balances fit quality against symbolic complexity. Cosmological predictions
are obtained from the normalized dark-energy density, while the structural
contribution to the description length is evaluated using the original
ESR expression tree. Our goals are to test whether the search recovers
familiar cosmological forms within its prescribed functional space and to
identify alternative, compact expressions that remain observationally
viable. In this way, we investigate how strongly current data constrain
the functional form of dark-energy evolution without restricting the
analysis to conventional parameterizations.

The paper is organized as follows. Section~\ref{sec:framework} presents
the cosmological framework and observables. Section~\ref{sec:methodology}
describes the ESR methodology, physical filters, parameter fitting, and
MDL-based selection. Section~\ref{sec:data} describes the observational
datasets, and Section~\ref{sec:results} presents our results. We discuss
their implications in Section~\ref{sec:discussion} and conclude in
Section~\ref{sec:conclusion}.


\section{Cosmological framework}\label{sec:framework}
To investigate the cosmological implications of the reconstructed dark-energy density, we first establish its relation to the background expansion history. We then introduce the cosmological observables and the likelihood function used to constrain the candidate expressions.
\subsection{Dark-energy density evolution}

We assume a spatially flat Friedmann–
Lemaître–Robertson–Walker (FLRW) universe  whose late-time expansion is described by matter and dark energy
alone, neglecting radiation.
The normalized Hubble expansion rate can then be written as
\begin{equation}
    E^2(z)\equiv\Omega_{\rm m,0}(1+z)^3+
    (1-\Omega_{\rm m,0})X(z),
    \label{eq:friedmann}
\end{equation}
where $E(z)\equiv H(z)/H_0$ is the dimensionless Hubble expansion rate, with $H_0=H(0)$ denoting the present-day Hubble parameter, $\Omega_{\rm m,0}$ is the present-day matter-density parameter and $\Omega_{\rm DE,0} = 1 - \Omega_{\rm m,0}$ is the present-day dark energy density parameter.

We define the normalized dark-energy density as
\begin{equation}
X(z)\equiv\frac{\rho_{\rm DE}(z)}{\rho_{\rm DE,0}} \,,
\label{eq:xdef}
\end{equation}
with $\rho_{\rm DE}(z)$ denoting the dark-energy density at redshift $z$ and $\rho_{\rm DE,0}\equiv\rho_{\rm DE}(0)$ its present-day value. By construction, $X(0)=1$.

For a cosmological constant, $X(z)=1$ at all redshifts. The equation-of-state parameter is related to the density evolution by 
\begin{equation}
    w(z)=-1+\frac{1+z}{3}\frac{d\ln X(z)}{dz} \,.
    \label{eq:wfromx}
\end{equation}

The smearing effect mentioned earlier can also be verified from the above equations. Clearly, working directly with $X(z)$ avoids introducing additional integrations between the reconstructed quantity and the expansion history \citep{Wang2001,Wang2006}. 


\subsection{Cosmological observables and likelihood}\label{sec:distances}

The late-time observables that constrain a candidate $X(z)$ are the BAO distances, the Type Ia supernova and a compressed CMB constraint. Each is mapped to a $\chi^2$ contribution, and the three are combined in a joint likelihood. The datasets are described in Section~\ref{sec:data}.

\subsubsection{BAO observables}
The analysis uses BAO measurements of the transverse comoving distance and Hubble distance, both rescaled by the sound horizon at the baryon-drag epoch ($r_d$)
\begin{equation}
    \tilde D_{\rm M}(z)\equiv\frac{D_{\rm M}(z)}{r_{\rm d}},
    \qquad
    \tilde D_{\rm H}(z)\equiv\frac{D_{\rm H}(z)}{r_{\rm d}}.
\end{equation}

For a spatially flat universe, the transverse comoving distance is given by $D_M(z)=(c/H_0)\int_0^z dz'/E(z')$, while the Hubble distance is defined as $D_H(z)=c/H(z)$. Since $H_0$ and $r_d$ enter the BAO observables through their product, we define
\begin{equation}
    \gamma \equiv \Omega_m \left( \frac{H_0 r_d}{c} \right)^2 \,, \quad q \equiv \frac{H_0 r_d}{c} = \sqrt{\frac{\gamma}{\Omega_m}} \,.
\end{equation}

The rescaled distances are then
\begin{equation}
    \begin{aligned}
    \tilde D_{\rm H}(z) &= \frac{1}{qE(z)}, \\
    \tilde D_{\rm M}(z) &= \int_0^z \tilde{D}_H(z') \, dz' \,.
    \end{aligned}
\end{equation}

The BAO data enter through,
\begin{equation}
    \chi^2_{\rm BAO}=\sum_k\Delta_k^T C_k^{-1}\Delta_k,
\label{eq:chi2bao}
\end{equation}
where $\boldsymbol{\Delta}_k$ is the data-minus-model vector of
$(\tilde D_{\rm M},\tilde D_{\rm H})$ in bin $k$ and $C_k$ is its $2\times2$
covariance (Sec.~\ref{sec:BAO}).


\subsubsection{Type Ia supernovae}
The Type Ia supernova analysis can be expressed in terms of the distance modulus,
\begin{equation}
    \mu_B=m_B-M_B
    =5\log_{10}\left(\frac{d_L}{\mathrm{Mpc}}\right)+25,
    \end{equation}
where $d_L(z)=(1+z)D_{\rm M}(z)$. We use the standardized apparent magnitudes rather than externally calibrated distance moduli. The formulation combines the supernova magnitude information with the BAO distance scale through the rescaled distance
\begin{equation}
    d_M(z)=\exp[b(20+M_B)]D_M(z),
    \quad b=\frac{\ln 10}{5},
\end{equation}
and the parameter
\begin{equation}
    \zeta=r_{\rm d}\exp[b(20+M_B)] \,,
\end{equation}
so that $d_M(z) = \zeta \tilde{D}_M (z)$. For a given candidate $X(z)$, the predicted apparent magnitude can be expressed as
\begin{equation}
    m_B^{\rm model}(z)=
    5\log_{10}\left[
    \frac{(1+z_{\rm hel})\zeta\widetilde D_M(z_{\rm HD})}
    {\mathrm{Mpc}}
    \right]+5 \,,
    \label{eq:mbmodel}
\end{equation}
where $z_{\rm HD}$ is the Hubble-diagram redshift and $z_{\rm hel}$ is the heliocentric one. 

Following \citet{Dinda2025}, $M_B$ and $r_d$ are absorbed into a single nuisance parameter $\zeta$, so no external absolute calibration is required. The supernova contribution is,
\begin{equation}
    \chi^2_{\rm SN} = \Delta \boldsymbol{m}^T C_{\rm SN}^{-1} \Delta \boldsymbol{m} \,,
    \label{eq:chi2sn}
\end{equation}
with $\Delta\boldsymbol{m}_i = m^{\rm obs}_{B,i} - m_B^{\rm model}(z_i)$ and $C_{\rm SN}$ is the Pantheon+ statistical$+$systematic covariance of the retained supernovae.


\subsubsection{Joint Likelihood}\label{sec:joint-like}

Let $\boldsymbol{\psi}=(\boldsymbol{\theta}_{\rm ESR},\Omega_{\rm m,0},\gamma,\zeta)$
denote the fitted parameters of a candidate. For Gaussian errors, each dataset
contributes $-\ln\mathcal{L}_i=\chi^2_i/2$ up to an additive constant independent
of $\boldsymbol{\psi}$. Treating the SN, BAO and CMB information as statistically
independent (we neglect correlations between BAO bins and between datasets), the
joint likelihood is the product of the individual ones and

\begin{equation}
    -2 \ln \mathcal{L}_{\rm tot} = \chi^2_{\rm SN} + \chi^2_{\rm BAO} + \chi^2_{\gamma}, 
    \label{eq:likelihood}
\end{equation}
where
\begin{equation}
    \chi^2_{\gamma} = \frac{(\gamma - \gamma_{\rm CMB})^2}{\sigma_{\gamma}^2} \,.
\end{equation}

The last term is not the full CMB likelihood but a Gaussian constraint on $\gamma$ compressing the Planck information, following \citet{Dinda2025} (Section~\ref{sec:CMB_prior}).  
The remaining parameters have uniform priors within the
bounds of Sec.~\ref{sec:ParFit}, which limit the search range but add no term.
Minimizing $-\ln\mathcal{L}_{\rm tot}$ gives the best-fit point and this quantity is the Residual term of the MDL.


\section{Methodology}\label{sec:methodology}

Symbolic Regression is a subfield of machine learning focused on inferring mathematical expressions from data \citep{Koza1994, SchmidtLipson2009, MakkeChawla2024}. Unlike usual parametric regression, where the functional form is fixed beforehand and only the parameters are optimized, SR includes the functional form itself in the search space: the structure of the expression (which operators compose it and how they combine) and the values of its numerical parameters are optimized simultaneously \citep{Bartlett_ESR}.

\subsection{ESR search space}

In ESR, the search is exhaustive within a predefined operator basis and equation complexity \citep{Bartlett_ESR}. Exhaustiveness refers to the enumeration of the admissible symbolic structures within the prescribed search space; it does not imply exhaustiveness over all mathematically possible functional forms. An important aspect of the ESR approach is the definition of symbolic complexity. We define the complexity $k$ of a mathematical expression as the total number of nodes in its tree representation, including operators, variables, and free parameters.  

The ESR search space is constructed from a predefined set of terminal symbols and mathematical operators. In this analysis, we adopt the following symbolic basis:
\begin{equation}
\{x,\ a,\ {\rm inv},\ {\rm exp},\ {\rm log},\ +,\ -,\ \times,\ /,\ {\rm pow}\},
\label{eq:n}
\end{equation}
where $x$ is the independent variable and $a$ denotes a free numerical parameter. The operators ${\rm inv}$, ${\rm exp}$, and ${\rm log}$ represent the reciprocal, exponential, and natural logarithm functions, respectively, while ${\rm pow}$ denotes exponentiation. Unary, binary, and nullary operators are combined recursively while enforcing tree arity and total complexity. Because distinct expression trees can represent equivalent functions, the implementation removes redundancy first through structural signatures and then through symbolic canonicalization with SymPy \citep{Meurer2017}. These steps reduce repeated fitting of structurally or algebraically equivalent expressions.


\subsection{Tree reconstruction and normalization}

Each generated tree is represented by $f_{\rm ESR}(x, \theta)$, where $x=1+z$, and is mapped to a candidate dark energy density evolution through
\begin{equation}
X(z)=\frac{f_{\rm ESR}(1+z;\theta)}{f_{\rm ESR}(1;\theta)},
\label{eq:normalization}
\end{equation}
which enforces $X(0)=1$ whenever the normalization is well defined. Candidates are rejected at this stage if  $f_{\mathrm{ESR}}(1;\boldsymbol{\theta})$ is not finite or has modulus smaller than $10^{-12}$, avoiding a numerically singular normalization.

Normalization has an important consequence for model comparison. Any overall multiplicative factor in \(f_{\rm ESR}(x,\boldsymbol{\theta}_{\rm ESR})\) that is independent of \(x\) cancels between the numerator and denominator of Equation~\ref{eq:normalization}, provided that the factor is nonzero. Consequently, structurally different ESR trees can map onto the same physical evolution \(X(z)\), including the constant solution \(X(z)=1\). The normalized function is used for cosmological predictions, whereas the original tree is retained for the structural part of the MDL score.


\subsection{Physical filtering \label{sec:PhysFilt}}

The normalized candidates are subjected to physical admissibility conditions motivated by the standard sequence of component-dominance epochs. First, the dark-energy density must remain positive, $X(z)>0$, over the prescribed evaluation domain. Second, it must be subdominant to matter and radiation over the corresponding reference epochs. We define the candidate-to-matter density ratio as
\begin{equation}
    R_m^{\rm ref}(z)_\equiv
    \frac{\rho_{\rm DE}(z)}{\rho_{\rm m}(z)}
    =
    \frac{\Omega_{\rm DE,0}^{\rm ref}}{\Omega_{\rm m,0}^{\rm ref}}
    \frac{X(z)}{(1+z)^3}\,,
    \label{eq:condition-m}
\end{equation}
we require dark energy to remain subdominant to matter over the matter-dominated reference interval,
\begin{equation}
    R_m^{\rm ref}(z) < 1 \,, \quad z_{\rm DE-m}^{\rm ref} < z \leq z_{\rm eq, r-m} \,,
\end{equation}
where the condition is tested on the prescribed interior sampling grid. The radiation-to-matter equality redshift is defined from the reference densities as
\begin{equation}
    z_{\rm eq, r-m} = \frac{\Omega^{\rm ref}_{\rm m,0}}{\Omega_{r,0}^{\rm ref}} -1\,.
\end{equation}

Similarly, the dark-energy-to-radiation ratio is
\begin{equation}
    R_{r}^{\rm ref}(z) \equiv \frac{\rho_{\rm DE}(z)}{\rho_r(z)} = \frac{\Omega_{\rm DE,0}^{\rm ref}}{\Omega_{r,0}^{\rm ref}} \frac{X(z)}{(1+z)^4} \,,
    \label{eq:condition-r}
\end{equation}
and the radiation-dominance requirement is
\begin{equation}
    R_r^{\rm ref}(z) < 1 \,, \quad z_{\rm eq, r-m} \leq z \leq z_{\rm BBN} \,.
\end{equation}

We adopt $z_{\rm BBN}=10^8$ as an operational upper redshift cutoff for the filter. This cutoff defines how far the expression is evaluated; it does not constitute a full calculation of the primordial nucleosynthesis history. In the reference backgrounds including radiation, spatial flatness gives $\Omega_{\rm DE,0}^{\rm ref}=1-\Omega_{\rm m,0}^{\rm ref}-\Omega_{\rm r,0}^{\rm ref}$.

The transition redshift $z_{\rm DE-m}^{\rm ref}$ is obtained from equality between matter and dark-energy densities in the chosen reference background, namely
\begin{equation}
\Omega_{\rm DE,0}^{\rm ref}X_{\rm ref}(z_{\rm DE-m}^{\rm ref})
=\Omega_{\rm m,0}^{\rm ref}(1+z_{\rm DE-m}^{\rm ref})^3,
\label{eq:zdem}
\end{equation}
where $X_{\rm ref}(z)$ is the reference model's normalized dark-energy evolution. For $\Lambda$CDM, $X_{\rm ref}(z)=1$; for a CPL reference, $X_{\rm ref}$ is the corresponding CPL density evolution. Equation~\ref{eq:zdem} is solved for each reference parameter set rather than applying the $\Lambda$CDM expression to a dynamical background.

\begin{table*}[hbt]
\centering
\footnotesize
\caption{Central parameters supplied to the physical-filter reference backgrounds. $H_0^{\rm ref}$ is in units of $\mathrm{km\,s^{-1}\,Mpc^{-1}}$.}
\label{tab:background-values}
\resizebox{\textwidth}{!}{
\begin{tabular}{lcccccccc}
\toprule
Model & $\Omega_{m,0}^{\rm ref}$ & $H_0^{\rm ref}$ & $\Omega_{r,0}^{\rm ref}$ & $\Omega_{\mathrm{DE},0}^{\rm ref}$ & $w_0$ & $w_a$ & $z_{\mathrm{DE-m}}^{\mathrm{ref}}$ & $z_{\mathrm{eq,r-m}}^{\rm ref}$ \\
\midrule
Planck 2018 base-$\Lambda$CDM & $0.3150$ & $67.40$ & $9.2064\times10^{-5}$ & $0.6849$ & $-1.00$ & $0.00$ & $0.29551$ & $3420.55$ \\
DESI DR2 + CMB ($\Lambda$CDM) & $0.3027$ & $68.17$ & $8.9996\times10^{-5}$ & $0.6972$ & $-1.00$ & $0.00$ & $0.32064$ & $3362.50$ \\
DESI DR2 + CMB ($w_0w_a$CDM) & $0.3530$  & $63.60$  & \(1.0339\times10^{-4}\) & $0.6469$ & $-0.42$ & $-1.75$ & $0.3568$ & $3413.14$ \\
\bottomrule
\end{tabular}
}
\par\smallskip
{\footnotesize Sources: Planck 2018 base-$\Lambda$CDM for the first row and DESI DR2 Results II, Table V, for the DESI rows \citep{Planck2018, DESI2025}. These reference parameters are used in physical filtering and are distinct from the cosmological parameters optimized in the SN+BAO+CMB likelihood. The exact radiation densities and derived transition redshifts are computed by the filtering utility.}
\end{table*}

The conditions in Equations~\ref{eq:condition-m} and \ref{eq:condition-r} require the candidate dark-energy contribution to remain subdominant to the component expected to dominate each reference interval.

We tested the physical conditions for three fixed reference backgrounds, listed in Table~\ref{tab:background-values}. For the DESI DR2+CMB $w_0w_a$CDM case, we used the marginal posterior means. The reference evolution is used only to define the epochs against which subdominance is tested; it does not impose a CPL form on the reconstructed candidate $X(z)$ or on its observational fit.

\vspace{0.2cm}

The physical admissibility test is sampled over the parameter space. For a candidate with $n_\theta$ free parameters, $N_{\rm try}=5$ parameter vectors are drawn from
\begin{equation}
    \theta_j^{\rm try}\sim \mathcal{U}[-5,5]^{n_\theta}
    \qquad j=1,\ldots,N_{\rm try}.
\label{eq:filter-sampling}
\end{equation}

A candidate is retained if at least one of the five sampled parameter vectors satisfies all implemented physical and numerical conditions. The resulting expressions passing these tests should therefore be interpreted as passing a finite sampling test rather than as being proven physical for every point in their parameter space.

After parameter optimization, the physical filter is re-evaluated using the fitted values of the free parameters (Section~\ref{sec:ParFit}). The transition redshifts are recalculated accordingly. 


\subsection{Parameter fitting}\label{sec:ParFit}

Each surviving candidate is fitted by minimizing the negative log-likelihood corresponding to Equation~\ref{eq:likelihood}. The ESR parameters and the cosmological sector $(\Omega_{m,0}, \gamma, \zeta)$ are optimized simultaneously. In our scaled-likelihood implementation, the cosmological variables are represented numerically by $(\Omega_{m,0}, \gamma_{\rm scaled}, \zeta_{\rm scaled})$ and mapped to the physical $\gamma$ and $\zeta$ entering the observables. The likelihood defines $\Omega_{m,0} \in [0.10, 0.60]$, $\gamma_{\rm scaled} \in [0.50, 1.50]$, and $\zeta_{\rm scaled} \in [0.50, 1.50]$; the base-line ESR bounds are $[-5,5]$, with wider ESR intervals used for the sensitivity test. The best-fit objective is evaluated at the full fitted vector $\hat{\boldsymbol{\psi}}=(\hat{\Omega}_{\rm m},\hat\gamma,\hat\zeta,\hat{\boldsymbol{\theta}}_{\rm ESR})$.


\subsection{MDL and parameter identifiability}

Likelihood minimization alone would favor increasingly flexible expressions whenever the additional complexity improves the fit. To compare candidates across different functional complexities, we adopt the construction of~\citep{Bartlett_ESR} for the symbolic component of the model, treating the cosmological nuisance sector as conditional on the joint observational analysis. Accordingly, only the parameters belonging to the ESR expression contribute to the parameter-description term, while $\Omega_m$, $\gamma$ and $\zeta$ are optimized as part of the likelihood but are not included in the symbolic MDL penalty. In the adopted formulation,
\begin{equation}
    \begin{aligned} 
        {\rm MDL}={}& -\ln\mathcal{L}(\hat{\boldsymbol{\psi}}) +k_{\rm ESR}\ln N_{\rm op} -\frac{p_{\rm ESR}}{2}\ln3\\ &+\sum_j\ln c_j +\sum_{i=1}^{p_{\rm ESR}} \left[ \frac12\ln I_{ii} +\ln|\hat{\theta}_i| \right] \,,
    \end{aligned}    
\label{eq:mdl}
\end{equation}
where $k_{\rm ESR}$ is the number of nodes in the original tree, $p_{\rm ESR}$ is the number of identifiable ESR parameters after the reduction procedure, $c_j$ are explicitly tracked simplification constants, and $I_{ii}$ are diagonal elements of the observed Fisher-information matrix in the ESR-parameter subspace. Here $N_{\rm op}=10$ is the number of symbols in the operator basis in Equation~\ref{eq:n}; it is not the number of observational data points. The terms are summarized in Table~\ref{tab:MDL} 

\begin{table*}[hbt]
\centering
\footnotesize
\caption{Terms entering the conditional symbolic MDL score.}
\label{tab:MDL}
\begin{tabular}{lll}
\toprule
Term & Interpretation & Implementation \\
\midrule
${\rm Residual}$ & Data fit & Joint SN+BAO+CMB likelihood \\
${\rm Function}$ & Symbolic structure & $k_{\rm ESR}\ln N_{\rm op}$ \\
${\rm Parameter}$ & Identifiable ESR parameters
    & Fisher-information and Rissanen terms \\
${\rm Constant}$ & Simplification constants
    & Explicitly tracked constants only \\
\bottomrule
\end{tabular}
\end{table*}

Because the physical evolution is defined through the normalized ratio in Equation~\ref{eq:normalization}, parameter directions that cancel identically in that ratio do not contribute to the effective parameter count. Their structural nodes nevertheless remain part of the original ESR representation and contribute to $k_{\rm ESR}$. Such exact cancellations are distinguished from parameters judged non-identifiable by the Fisher-based criterion.

For the remaining ESR parameters, identifiability is assessed from the diagonal observed Fisher information in the ESR subspace. In our implementation, $I_{ii}$ is the second derivative of the negative log-likelihood objective with respect to $\theta_i$, evaluated at the joint best-fit vector while all other fitted coordinates, including the cosmological nuisance parameters, are held fixed:
\begin{equation}
I_{ii}=\left.\frac{\partial^2[-\ln\mathcal{L}(\boldsymbol{\psi})]}{\partial\theta_i^2}\right|_{\boldsymbol{\psi}=\hat{\boldsymbol{\psi}}}.
\end{equation}

The characteristic parameter resolution is
\begin{equation}
\Delta_i=\sqrt{\frac{12}{I_{ii}}}.
\end{equation}

Parameters satisfying
\begin{equation}
\frac{|\hat\theta_i|}{\Delta_i}<1
\end{equation}
are treated as non-identifiable at the adopted precision, removed from the expression, and followed by a refit of the full likelihood, with the cosmological nuisance parameters re-optimized. The Fisher-based test is a practical diagonal criterion and does not by itself capture all correlations among parameters. The final comparison is expressed as
\begin{equation}
\Delta\mathrm{MDL}_i=\mathrm{MDL}_i-\mathrm{MDL}_{\min}.
\end{equation}


\subsection{Functional classification}\label{sec:family}

To identify known cosmological parameterizations within the ESR output, final expressions are compared with three reference families: $\Lambda$CDM, Chevallier–Polarski–Linder (CPL)~\citep{Chevallier2001, Linder2003}, and Barboza-Alcaniz (BA)~\citep{Barboza2008}. Classification is performed on the final reconstructed \(X_{\rm final}(z)\), after any parameter reduction, using the logarithmic derivative $G(z)\equiv d\ln X/dz$. 

For $\Lambda$CDM, the dark-energy density is constant, $X_{\Lambda{\rm CDM}}(z)=1$, implying $G_{\Lambda{\rm CDM}}(z)=0$.

For CPL, $w(z)=w_0+w_a z/(1+z)$, which gives
\begin{equation}
\begin{aligned}
X_{\rm CPL}(z) &= (1+z)^{3(1+w_0+w_a)}
\exp\left[-3w_a\frac{z}{1+z}\right],\\
G_{\rm CPL}(z) &= \frac{3(1+w_0+w_a)}{1+z}
-\frac{3w_a}{(1+z)^2}.
\end{aligned}
\end{equation}

Similarly, for BA, $w(z)=w_0+w_a z(1+z)/(1+z^2)$, we obtain
\begin{equation}
\begin{aligned}
X_{\rm BA}(z) &= (1+z)^{3(1+w_0)}
(1+z^2)^{3w_a/2},\\
G_{\rm BA}(z) &= \frac{3(1+w_0)}{1+z}
+\frac{3w_a z}{1+z^2}.
\end{aligned}
\end{equation}

Writing $x=1+z$, the implementation classifies both the generated normalized expression and the final retained expression testing the symbolic equivalence rather than a numerical comparison.  The CPL and BA parameters are inferred algebraically from the symbolic form, and $\Lambda$CDM is checked first because it is a special case of both families. Expressions for which the symbolic equivalence cannot be established are labeled OTHER. 


\section{Data and Analysis}\label{sec:data}

The analysis combines three sources of cosmological information: Type Ia supernovae, DESI DR2 BAO measurements, and a compressed CMB prior. Each candidate $X(z)$ is inserted into Equation~\ref{eq:friedmann}, propagated into the distances defined in Section~\ref{sec:distances}, and evaluated against the selected supernova covariance and the block-diagonal BAO covariance approximation described below.

\subsection{Type Ia supernovae}\label{sec:sn-obs}

We use the Pantheon+ compilation \citep{Scolnic2022}, which contains 1701 light curves corresponding to 1550 distinct supernovae. We use standardized apparent magnitudes $m_{B,\rm corr}$ rather than SH0ES-calibrated distance moduli. We retain observations with $z_{\rm HD}\geq0.01$ to limit the impact of peculiar velocities and remove all Cepheid-host calibrators. Because no absolute-distance calibration is imposed, $M_B$ is not fixed and is absorbed into $\zeta$; the supernova data do not independently constrain $H_0$ in this formulation. 


\subsection{Baryon acoustic oscillations}\label{sec:BAO}

We use six DESI DR2 BAO bins \citep{DESI2025}, listed in Table~\ref{tab:bao}. Each bin provides $\widetilde D_{\rm M}$ and $\widetilde D_{\rm H}$ and their correlation coefficient $r_{\rm M,H}$. The adopted covariance is block diagonal, with one $2\times2$ block per bin:
\begin{equation}
    C_k=\begin{pmatrix}
    \sigma_{{\rm M},k}^2 & r_k\sigma_{{\rm M},k}\sigma_{{\rm H},k}\\
    r_k\sigma_{{\rm M},k}\sigma_{{\rm H},k} & \sigma_{{\rm H},k}^2
    \end{pmatrix},
\end{equation}
Correlations between different bins are neglected in the present analysis. 

\begin{table}[t]
\centering
\footnotesize
\setlength{\tabcolsep}{2.5pt}
\caption{DESI DR2 BAO measurements used in this work.}
\label{tab:bao}
\begin{tabular}{lcccc}
\toprule
Tracer & $z_{\rm eff}$ & $\widetilde D_{\rm M}$ & $\widetilde D_{\rm H}$ & $r_{\rm M,H}$\\
\midrule
LRG1       & 0.510 & $13.588\pm0.167$ & $21.863\pm0.425$ & $-0.459$\\
LRG2       & 0.706 & $17.351\pm0.177$ & $19.455\pm0.330$ & $-0.404$\\
LRG3+ELG1  & 0.934 & $21.576\pm0.152$ & $17.641\pm0.193$ & $-0.416$\\
ELG2       & 1.321 & $27.601\pm0.318$ & $14.176\pm0.221$ & $-0.434$\\
QSO        & 1.484 & $30.512\pm0.760$ & $12.817\pm0.516$ & $-0.500$\\
Ly$\alpha$ & 2.330 & $38.988\pm0.531$ & $8.632\pm0.101$  & $-0.431$\\
\bottomrule
\end{tabular}
\end{table}


\subsection{Compressed CMB prior\label{sec:CMB_prior}}

Following \citet{Dinda2025}, CMB information enters as a Gaussian prior on $\gamma$, derived from Planck 2018 base-$\Lambda$CDM information about the early-time cosmological model. We adopt
\begin{equation}
\gamma_{\rm CMB}=(343.8\pm1.6)\times10^{-6}.
\end{equation}
This compressed prior is not the full CMB likelihood. It assumes standard pre-recombination physics and is used here to constrain the parameter combination relevant to the BAO distance scale. In particular, since $\gamma = \Omega_m (H_0 r_d/c)^2$, the inferred constraint depends on the sound horizon $r_d$, which is sensitive to the pre-recombination expansion history. Dark-energy models that significantly modify the early-Universe evolution may therefore alter both $r_d$ and the corresponding CMB constraint on $\gamma$. Its use should thus be understood as conditional on the assumed early-Universe physics, rather than as a model-independent replacement for the full CMB likelihood.


\section{Results \label{sec:results}}

We begin by examining the reduction of the symbolic search space and the yield of the subsequent fitting procedure. The ESR generator produced 361,432 candidate expressions across complexities 1--8. After applying the physical admissibility condition, 179,331 expressions were retained for parameter estimation. We then examine the resulting MDL ranking, the reconstructed dark energy density histories, and the corresponding cosmological distance observables. Finally, we discuss the implication of these results for the current debate on dynamical dark energy.

\subsection{Search-space reduction and fitting yield}\label{sec:fit-yield}

The physical filter admitted 179,331 candidate trees across complexities 1--8 for the DESI DR2+CMB $w_0w_a$CDM reference background. The fitting stage saved 141,677 results, corresponding to a fraction of 79.0\% of the filtered input catalogue (the per-complexity accounting is reported in Appendix~\ref{app:fit-completion}). The consolidated post-processing then reduced these results to 35,083 symbolically distinct normalized expressions and, after numerical curve-equivalence grouping, 7,539 final representatives. All subsequent ranks in this section refer to this final numerical-curve catalogue.

The parameter-reduction procedure was not needed for most final curve representatives. In the 7,539 final catalogue, 599 representatives (7.95\%) were associated with parameter removal and refitting, whereas
6,940 (92.05\%) retained their parameterization without such a reduction.

\subsection{MDL ranking and leading symbolic expressions}\label{sec:mdl-ranking}

The global minimum of the adopted MDL score is the constant-density solution. It is represented by the complexity 1 tree $f_{\rm ESR}(x,\theta)=\theta_0$, for which the normalization in Equation~\ref{eq:normalization} removes the overall constant and gives $X_{\rm final}(z)=1$. The nominal tree parameter therefore does not remain in the normalized cosmological function and contributes no effective ESR-parameter penalty. Its total score is
${\rm MDL}=702.353$, the lowest value in the final catalogue. Thus, $\Lambda$CDM is recovered as the most economical description within the operator basis, complexity range, physical-filter prescription, data
combination, and conditional MDL convention adopted here.

The leading nonconstant reconstruction is the complexity 3 expression $f_{\rm ESR}(x,\theta)=\theta_0+x$, which is fitted at the upper boundary $\hat\theta_0=5$. At the fitted value, this gives
\begin{equation}
    X_{\rm final}(z) = \frac{\hat\theta_0 +1+z} {\hat\theta_0+1} = 1 + \frac{z}{6} \,,
\end{equation}
its score is ${\rm MDL}=708.428$, or $\Delta{\rm MDL}=6.075$ relative to the minimum. The next-ranked candidate is the complexity 4 tree $f_{\rm ESR}(x,\theta)=\exp(\theta_0/x)$, with $\hat\theta_0=-0.171786$, yielding
\begin{equation}
X_{\rm final}(z)=\exp\!\left[-\hat\theta_0\frac{z}{1+z}\right].
\label{eq:best-nonconstant-cpl}
\end{equation}

This expression belongs to the CPL family along the one-parameter relation $w_0+w_a=-1$, with $w_a=\hat\theta_0/3$ and $w_0=-1-\hat\theta_0/3$. Its score is ${\rm MDL}=708.589$. For the fitted value $\hat{\theta}_0=-0.171786$, we obtain $w_0\simeq-0.943$ and $w_a\simeq-0.057$, corresponding to $w(z)>-1$ for $z\geq0$. Thus, this reconstruction provides an example of dynamical dark energy that modestly improves the fit without requiring a crossing of the phantom divide. The distinction between goodness of fit and MDL ranking is particularly evident for these two candidates. The rank-2 linear-density reconstruction has a slightly larger residual contribution than $\Lambda$CDM (700.408 versus 700.051), corresponding to $\Delta\chi^2\simeq +0.71$. In contrast, the rank-3 exponential reconstruction improves the fit, with a residual contribution of 698.509 ($\Delta\chi^2\simeq -3.08$). Nevertheless, both expressions have higher total MDL scores than the constant-density solution, illustrating that an acceptable or improved observational fit does not necessarily imply a preference under the adopted complexity-penalized criterion.

Table~\ref{tab:mdl-top-ranked} lists the ESR equations with the lowest MDL for the dark-energy evolution $X(z)$. The individual terms of the MDL are represented as residual, function, parameter, and constant columns, and their sum gives the total score. The parameter term includes the adopted ESR parameter-description and Fisher/Rissanen contributions, while the constant term contains only explicitly tracked simplification constants. This decomposition allows us to examine how the balance between fit quality and descriptive complexity determines the final ranking. 

\begin{table*}[t]
\centering
\scriptsize
\setlength{\tabcolsep}{2.1pt}
\renewcommand{\arraystretch}{2.5}
\caption{Lowest-MDL expressions. The table shows ranks 1--10 and the 150th-ranked numerical curve
representative. Here $x=1+z$; $f_{\rm ESR}$ is the original symbolic tree,
whereas $X_{\rm final}$ is its normalized expression after the implemented
parameter-reduction procedure. All MDL components are rounded to three
decimal places.}
\label{tab:mdl-top-ranked}
\resizebox{\textwidth}{!}{%
\begin{tabular}{ccccccccc}
\toprule
Rank & $f_{\rm ESR}(x,\theta)$ & $X_{\rm final}(z)$ & Complexity
& Residual & Function & Parameter
& Constant & $ {\rm MDL}$ \\
\midrule
1 & $\theta_0$ & $1$ & 1 & 700.051 & 2.303 & 0.000 & 0.000 & 702.353 \\
2$^{\dagger}$ & $\theta_0+x$ & $\displaystyle\frac{\theta_0+1+z}{\theta_0+1}$
& 3 & 700.408 & 6.908 & 1.113 & 0.000 & 708.428 \\
3 & $\exp(\theta_0/x)$
& $\displaystyle\exp\!\left[-\theta_0\frac{z}{1+z}\right]$
& 4 & 698.509 & 9.210 & 0.869 & 0.000 & 708.589 \\
4$^{\dagger}$ & $\log|\theta_0+x|$
& $\displaystyle\frac{\ln|\theta_0+1+z|}{\ln|\theta_0+1|}$
& 4 & 699.176 & 9.210 & 0.819 & 0.000 & 709.205 \\
5$^{\dagger}$ & $\theta_0-1/x$
& $\displaystyle\frac{\theta_0(1+z)-1}{(1+z)(\theta_0-1)}$
& 4 & 698.687 & 9.210 & 1.443 & 0.000 & 709.340 \\
6 & $x^{1/\exp(x)}$
& $\displaystyle(1+z)^{\exp[-(1+z)]}$
& 5 & 698.001 & 11.513 & 0.000 & 0.000 & 709.514 \\
7 & $x^{\exp(\theta_0)}$
& $\displaystyle(1+z)^{\exp(\theta_0)}$
& 4 & 698.869 & 9.210 & 1.449 & 0.000 & 709.529 \\
8$^{\dagger}$ & $\theta_0-\log|x|$
& $\displaystyle1-\frac{\ln(1+z)}{\theta_0}$
& 4 & 699.195 & 9.210 & 1.205 & 0.000 & 709.610 \\
9 & $x+\exp(\theta_0)$
& $\displaystyle\frac{1+z+\exp(\theta_0)}{1+\exp(\theta_0)}$
& 4 & 699.290 & 9.210 & 1.260 & 0.000 & 709.760 \\
10$^{\dagger}$ & $x^{1/\theta_0}$
& $\displaystyle(1+z)^{1/\theta_0}$
& 4 & 699.456 & 9.210 & 1.307 & 0.000 & 709.973 \\
\multicolumn{9}{c}{$\vdots$} \\
150 & $\displaystyle\frac{1}{\theta_0+\log|x|/x}$
& $\displaystyle\frac{\theta_0(1+z)}{\theta_0(1+z)+\ln(1+z)}$
& 7 & 698.179 & 16.118 & 1.025 & 0.000 & 715.322 \\
\bottomrule
\end{tabular}%
}
\par\smallskip
\begin{minipage}{0.98\textwidth}
\scriptsize
\textit{Notes.} Residual = $- \ln \mathcal{L}_{\rm tot}(\hat{\theta})$; Function $= k_{\rm ESR}\ln N_{\rm op}$; Parameter$ = -\frac{p_{\rm ESR}}{2}\ln3+\sum_{i=1}^{p_{\rm ESR}} \left[ \frac12\ln I_{ii} +\ln|\hat{\theta}_i| \right] $; and Constant contains the tracked simplification-constant terms.
The fitted values of the one-parameter expressions, in rank order, are
$\hat\theta_0=5.000$ (rank 2), $-0.171786$ (rank 3), $5.000$ (rank 4),
$5.000$ (rank 5), $-2.140218$ (rank 7), $-5.000$ (rank 8), $2.539109$
(rank 9), $5.000$ (rank 10), and $-4.334938$ (rank 150). The dagger marks
fits at an imposed ESR-parameter boundary, $\hat\theta_0=\pm5$. The rank-1
parameter is removed because it cancels identically in the normalized
function. The rank-150 representative has $\Delta{\rm MDL}=12.969$.
\end{minipage}
\end{table*}

The best nonconstant candidate at rank 2 and several other highly ranked expressions reach the imposed interval boundary. In particular 31 of the 150 selected representatives (20.7\%) have at least one retained ESR parameter at a fitting boundary. The separate fitting-bound sensitivity test in Appendix~\ref{app:range-sensitivity} was performed for complexities 4 and 5 and does not establish bound-independence for every complexity represented in Table~\ref{tab:mdl-top-ranked}. Therefore, the reported ranking is referred to the fitting interval adopted for the ESR parameters. 

Among the 150 lowest-MDL numerical curve representatives, the implemented symbolic classifier assigns 1 expression to $\Lambda$CDM, 6 to CPL, and 143 expressions with no family correspondence; no BA expression occurs in this subset.  
In this hierarchy, CPL also includes its $w_a=0$ subfamily.

\subsection{Selected dark-energy density reconstructions}\label{sec:X-reconstructions}

\begin{figure}[hbt]
\centering
\includegraphics[width=0.5\textwidth]{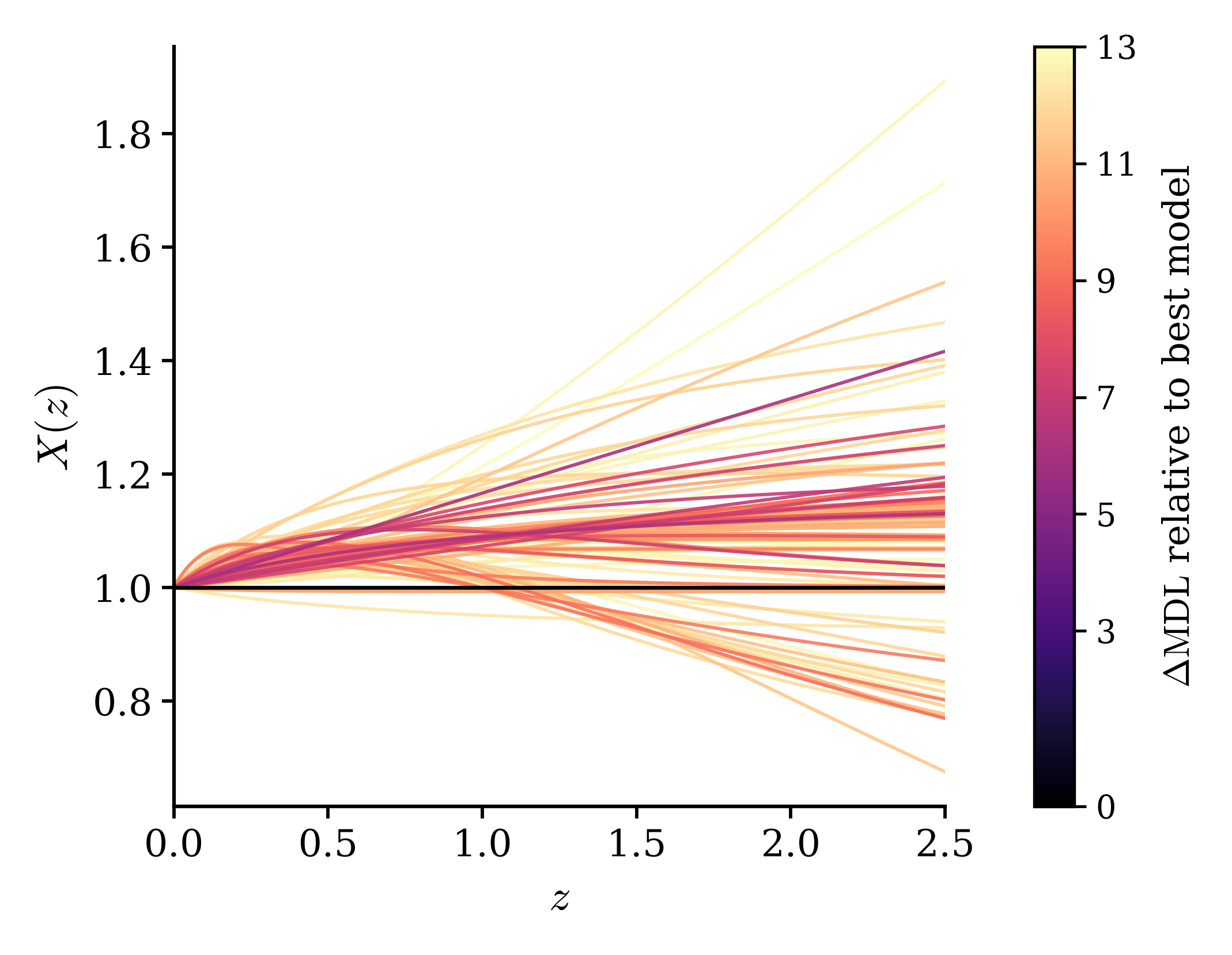}
\caption{Individual normalized dark-energy density reconstructions for the
150 lowest-MDL numerical curve representatives. The rank-150 curve is the upper endpoint
of the plotted sample, with $\Delta{\rm MDL}=12.969$.}
\label{fig:X-ranked-curves}
\end{figure}

Fig.~\ref{fig:X-ranked-curves} displays the normalized density histories for the 150 lowest-MDL candidates, colored by their difference relative to the minimum MDL. The high-ranked expressions span different functional shapes, although their redshift evolution remains relatively close to unity for much of the fitted interval. The wider ensemble of the 150 selected curves spans approximately $0.676 \lesssim X(2.5) \lesssim 1.893$, illustrating the diversity of dark-energy density histories represented among the 150 lowest-MDL candidates.

\subsection{Distance observables and functional degeneracy}\label{sec:distance-reconstructions}

\begin{figure}[hbt]
\centering
\includegraphics[width=0.5\textwidth]{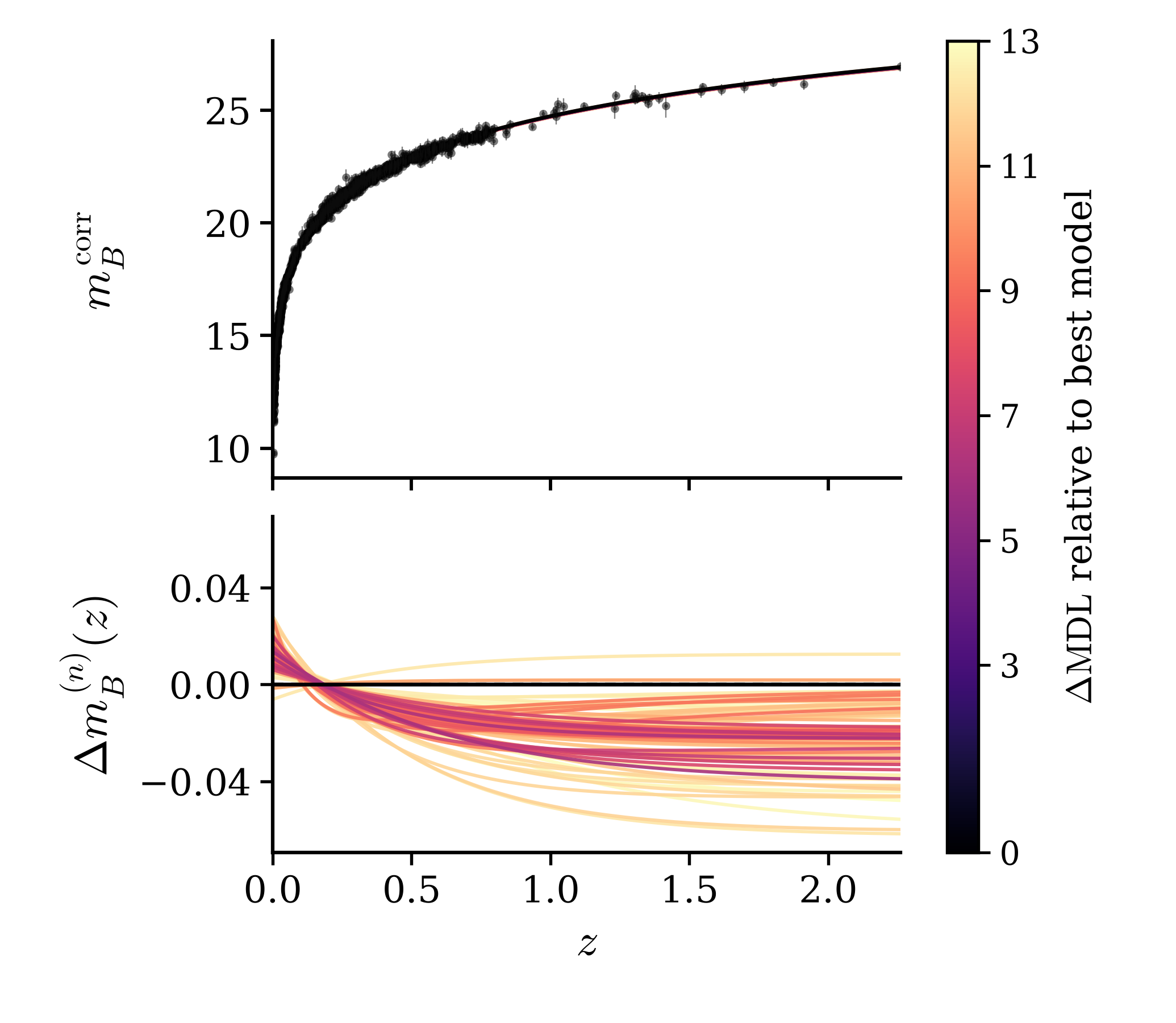}
\caption{In the top panel we present the standardized Type Ia supernova apparent-magnitude reconstruction
for the 150 lowest-MDL numerical curve representatives. Points and error
bars show the selected Pantheon+ magnitude sample used in the likelihood;
Cepheid-host calibrators are removed and no SH0ES absolute-magnitude
calibration is imposed. In the bottom we show the deviation from the minimum-MDL $\Lambda$CDM prediction. Curves are color-coded by $\Delta{\rm MDL}$.}
\label{fig:sn-reconstruction}
\end{figure}

\begin{figure*}[hbt]
\centering
\includegraphics[width=0.49\textwidth]{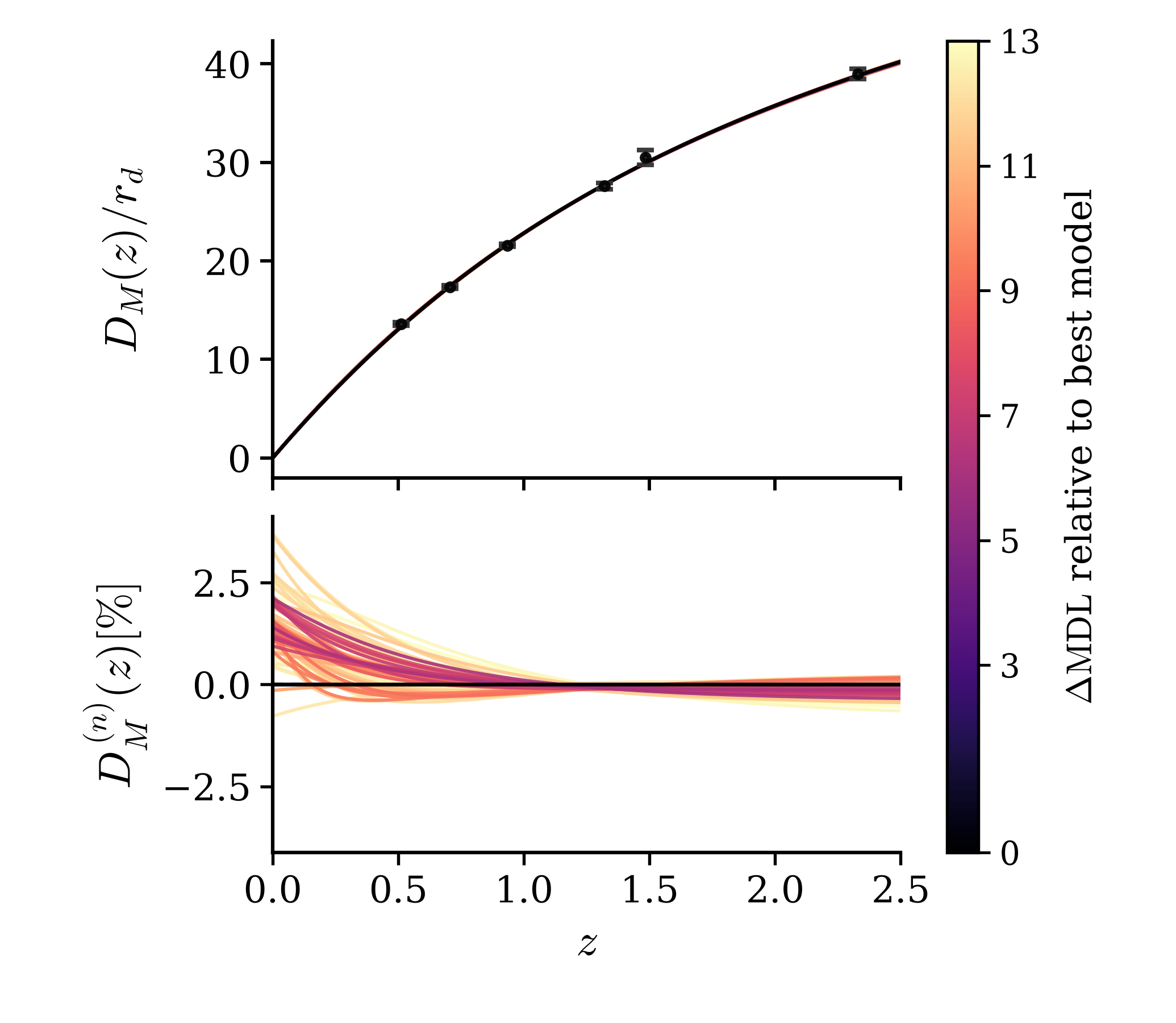}
\hfill
\includegraphics[width=0.49\textwidth]{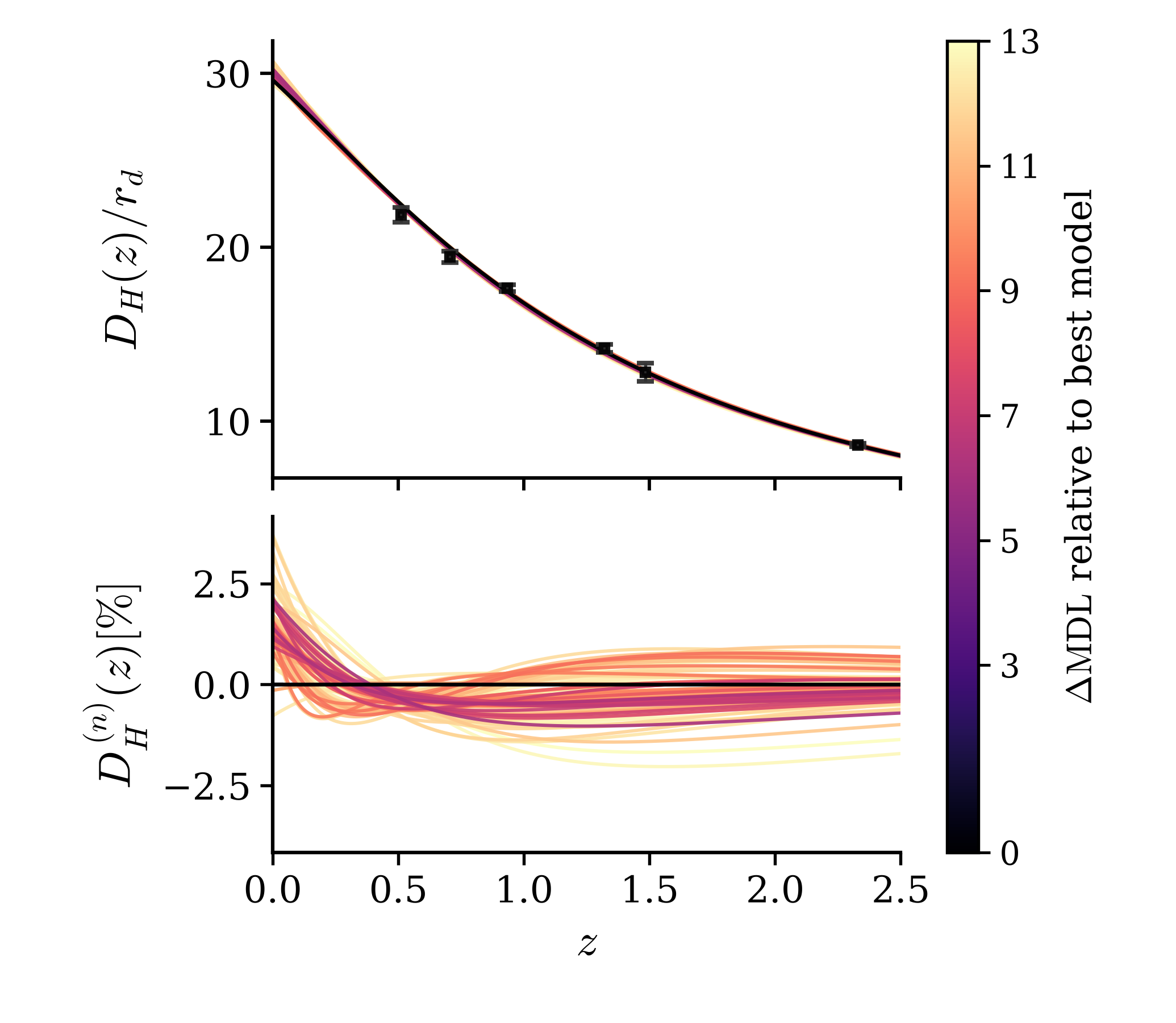}
\caption{In the top panel we present the transverse (left) and radial (right) BAO distance reconstructions for the 150 lowest-MDL numerical curve representatives.
The points and error bars show the six DESI DR2 measurements. 
In the bottom we show the percentage deviation from the minimum-MDL $\Lambda$CDM prediction,
$D_i^{(n)}(z)=100\,[\widetilde D_i^{(n)}-\widetilde D_i^{\Lambda{\rm CDM}}]/\widetilde D_i^{\Lambda{\rm CDM}}$,
$i=M,H$. Each curve is evaluated at its own best-fit $(\Omega_{\rm m,0},\gamma,\zeta)$
and colored by $\Delta{\rm MDL}$. }
\label{fig:bao-reconstruction}
\end{figure*}

The reconstructed distance observables are shown in Figs.~\ref{fig:sn-reconstruction} and~\ref{fig:bao-reconstruction}. The plotted curves correspond to the same 150 lowest-MDL numerical curve representatives, color-coded by their difference in MDL relative to the minimum. The supernova panel shows the standardized apparent-magnitude data used in the likelihood; the Cepheid-host calibrators are excluded and no external SH0ES absolute-distance calibration is imposed.

The predicted supernova magnitudes are closely clustered across the selected models. The same behavior is seen in the transverse BAO distance $\tilde D_{\rm M}$ and, with a somewhat larger dispersion, in the radial BAO distance $\tilde D_{\rm H}$. For example, near the highest-redshift Ly$\alpha$ BAO bin ($z\simeq2.33$), the 150-model envelope spans less than about 0.8\% in $\tilde D_{\rm M}$ and about 2.7\% in $\tilde D_{\rm H}$.

These narrow distance envelopes contrast with the broader spread in the reconstructed $X(z)$ histories. At $z\simeq2.33$, the selected density reconstructions span approximately $0.72$--$1.81$, while their predicted
integrated distance observables remain much more tightly clustered. This illustrates a functional degeneracy of the present distance-based reconstruction: distinct dark-energy density histories can yield very similar predictions for the observables used in the likelihood. The plotted ensemble is selected by MDL rank and is not a statistical confidence region.

To make the differences visible, the lower panels of Figs.~\ref{fig:sn-reconstruction}
and~\ref{fig:bao-reconstruction} show the deviation of each prediction from that of the
minimum-MDL candidate, $X(z)=1$ ($\Lambda$CDM), both evaluated at their own best fit:
$\Delta m_B^{(n)}(z)=m_B^{(n)}(z)-m_B^{\Lambda{\rm CDM}}(z)$ and $\Delta_i^{(n)}(z)$ as defined
in the caption, with $n$ labeling the 150 reconstructions.

\section{Discussion}\label{sec:discussion}

The principal result of this analysis is that the constant-density solution $X(z)=1$ provides the minimum-MDL reconstruction within the explored ESR catalogue. Several nonconstant expressions, including the linear-density model at rank 2 and the CPL subfamily at rank 3, reproduce the observational data with comparable goodness of fit. However, these models are not preferred by the adopted MDL criterion. While the rank-2 model provides a slightly worse fit than $\Lambda$CDM, the rank-3 reconstruction achieves a modest improvement that is outweighed by its additional descriptive complexity. Thus, the ability of nonconstant dark-energy histories to reproduce current observations should be distinguished from whether their additional functional complexity is justified under the adopted MDL criterion.

This result is consistent with model-independent reconstructions of the dark energy density. Using Gaussian processes with DESI~DR2 BAO and three SN samples, \citet{deSouza2026} found that $\Lambda$CDM, CPL and thawing quintessence all agree with the reconstructed density at the 95\% confidence level, although $\Lambda$CDM shows the largest discrepancy (with DESY5, at low redshift) and the reconstructed shape itself suggests phantom crossing; the authors conclude that it is premature to claim significant evidence for evolving dark energy. \citet{WangFreese2026}, using DESI~DR2 BAO and Planck distance priors, also find agreement with a constant density except for a $1$--$2\sigma$ deviation at $0.4\lesssim z\lesssim0.9$, and attribute the $3.1\sigma$ DESI preference to the assumed $w_0w_a$ parametrization. Closest to our approach, \citet{Sousa-Neto:2025gpj} applied symbolic regression (PySR) to reconstruct $w(z)$ from DESI~DR2 and SN data and found no deviation from $\Lambda$CDM within $3\sigma$. The criteria differ: a reconstruction asks whether $X=1$ lies within the allowed region, whereas the MDL ranking asks whether any expression in a finite basis improves the fit enough to pay for its complexity.

The significance of departures depends on the SN sample. \citet{Berti2025}, using DESI~DR1  BAO, CMB and SNe, report $2.4\sigma$ with DESY5 but $1.3\sigma$ with Pantheon+ (with phantom crossing in the reconstructed equation of state in all cases), and confirm that correcting the low-redshift magnitude offset of DESY5 removes the discrepancy. Our preference for $X=1$ should therefore be read as conditional on the Pantheon+ compilation; repeating the ranking with Union3 or DESY5 is a natural test. This illustrates the sensitivity of reconstructed dark-energy histories to the observational data combination and to possible systematic effects.

Our results should therefore be interpreted in the context
of the adopted Pantheon+ compilation, DESI DR2 BAO measurements, compressed Planck prior, and conditional MDL prescription. Rather than excluding dynamical dark energy, the present analysis indicates that a constant density is the most economical description within the
explored symbolic space, while a range of nonconstant histories continues to reproduce the observed distances. 

Some methodological choices delimit the scope of this result. The
physical filter uses first-order component-dominance conditions: dark energy
is required to remain subdominant to matter and radiation over the
corresponding reference epochs. This is an operational admissibility test,
not a complete early-dark-energy constraint or a full treatment of
pre-recombination physics. The compressed CMB prior is conditional on
standard pre-recombination assumptions, and candidates with a non-negligible
early dark-energy fraction are not independently tested against a full CMB
likelihood. A more stringent treatment of early-Universe constraints is
left for future work.

The parameter-identifiability procedure follows the practical diagonal
Fisher-information prescription adopted in the ESR methodology. It is a
conditional criterion and does not fully describe correlations among all
fitted parameters. In the final numerical-curve catalogue, however, only
599 of 7,539 representatives (7.95\%) are associated with parameter
removal and refitting; the remaining 92.05\% retain their parameterization.
The present rankings should be understood within this adopted first
implementation of the identifiability procedure.

The final ranking is also conditional on the finite symbolic search space,
the chosen independent variable $x=1+z$, the operator basis, the finite
parameter-sampling procedure used by the physical filter, and the fitting
bounds. Several leading nonconstant candidates reach a fitting boundary,
which is why the bound sensitivity results are reported separately.

A similar dependence on the adopted symbolic representation was investigated by \citet{Sousa_2024} in the context of inflationary potentials. Using ESR and MDL, they explored how different operator bases and structural priors affect the ranking of candidate expressions. Although their analysis concerns a different cosmological regime, it highlights the importance of interpreting our preference for $X(z)=1$ within the assumptions defining the symbolic search space.

These methodological choices, together with the adopted observational data and background assumptions, define the scope of the present analysis. Extending the study to alternative supernova compilations, additional BAO observables, full cross-bin covariance, and more detailed early-time physics provides natural directions for future work.

\section{Conclusion}\label{sec:conclusion}

We have investigated how strongly current cosmological observations constrain the functional form of dark-energy evolution without restricting the analysis to conventional phenomenological parameterizations. To this end, we developed an Exhaustive Symbolic Regression (ESR) framework to reconstruct the normalized dark-energy density, $X(z)$, combining exhaustive symbolic enumeration, physically motivated admissibility conditions, parameter estimation, and a conditional MDL criterion that balances goodness of fit against symbolic complexity. Using Pantheon+ Type Ia supernovae, DESI DR2 BAO measurements, and a compressed CMB prior, we examined whether this approach recovers familiar cosmological models and identifies alternative, compact descriptions of dynamical dark energy that remain observationally viable.

In particular, the incorporation of physically motivated admissibility plays a central role in restricting the symbolic search space. While symbolic regression can generate a large variety of mathematically valid expressions, not all of them necessarily correspond to physically acceptable dark energy histories. In our ESR generation procedure, 361,432 candidate expression trees were generated across the symbolic complexities 1--8. These expressions were subsequently subjected to physically motivated admissibility conditions, which
require positive dark-energy density and subdominance relative to matter and radiation over the corresponding reference epochs. The physical filter retained 179,331 expressions (49.62\% of the initial catalogue), rejecting approximately half of the generated candidates.

The constant-density solution $X(z)=1$, corresponding to $\Lambda$CDM, yields the minimum MDL among the final catalogue (MDL=702.353). Our results therefore favor a cosmological constant over dynamical dark energy as the most economical description within the explored symbolic search and physical assumptions. This preference does not arise from imposing a constant density as a prior functional form, in fact it emerges from the comparison of the generated expressions after parameter optimization and model selection. Although some nonconstant reconstructions achieve modest improvements in the likelihood, these improvements are insufficient to compensate for their additional descriptive complexity. In particular, the best-ranked nonconstant expression does not improve the fit relative to $\Lambda$CDM, whereas the third-ranked candidate achieves a modest improvement but remains disfavored by the adopted MDL criterion. Nevertheless, the existence of nonconstant reconstructions that reproduce the observed distances highlights the distinction between observational compatibility and preference under complexity-based model selection.

An additional result concerns the functional degeneracy among the reconstructed histories. The 150 lowest-MDL
representatives exhibit a substantially broader range of $X(z)$ than their corresponding supernova and BAO distance predictions would suggest. This indicates that distinct dark-energy histories can produce very similar integrated cosmological observables, highlighting the limited ability of current distance measurements to uniquely determine the functional form of $X(z)$.  

The conclusions are conditional on the selected operator basis, physical
filter, fitting bounds, observational data combination, and MDL prescription.
A more detailed early-Universe admissibility treatment and a fuller analysis
of parameter degeneracies are important extensions, but are beyond the
scope of the present first implementation.

\begin{acknowledgments}
AM acknowledges Coordenação de Aperfeiçoamento de Pessoal de Nível Superior - CAPES for financial support. ASN is supported by the National Observatory PCI-DB program. JA is supported by the Conselho Nacional de Desenvolvimento Científico e Tecnológico (CNPq) grants (No. 307683/2022-2 and 448158/2025-6) and Fundação de Amparo à Pesquisa do Estado do Rio de Janeiro (FAPERJ) grant No. 299312 (2023). The development of this work was aided by the National Observatory Data Center (CPDON).  
\end{acknowledgments}

\appendix
\restartappendixnumbering

\section{Sensitivity of the physical filtering to the sampling interval}\label{app:filter-sensitivity}

The physical filter evaluates each candidate expression at a finite number of trial parameter vectors drawn from a prescribed uniform interval. Consequently, the resulting survivor catalogue depends on the range adopted for this sampling. To assess this dependence, we repeated the filtering of the combined complexities 4 and 5 catalogue, which contains 1,280 expressions, drawing the trial vectors from $[-5,5]$, $[-10,10]$, and $[-20,20]$. The corresponding numbers of surviving expressions were 786, 777, and 767, respectively, as reported in Table~\ref{tab:filter-bound-sensitivity}.

\begin{table}[hbt]
\centering
\footnotesize
\caption{Number of C4--C5 expressions surviving the physical filter for
different ranges used to sample the trial ESR parameter vectors.}
\label{tab:filter-bound-sensitivity}
\begin{tabular}{lrrr}
\toprule
Trial-parameter interval & C4 survivors & C5 survivors & Total survivors \\
\midrule
$[-5,5]$ & 114 & 672 & 786 \\
$[-10,10]$ & 113 & 664 & 777 \\
$[-20,20]$ & 112 & 655 & 767 \\
\bottomrule
\end{tabular}
\end{table}

The number of surviving expressions decreases slightly as the sampling range
is widened. The difference between the narrowest and widest ranges is 19
expressions, corresponding to a reduction of approximately 2.4\% relative to
the $[-5,5]$ result. Thus, within the range of configurations tested, the
choice of sampling range has a modest effect on the size of the survivor
catalogue.

We adopt $[-5,5]$ as the default sampling range in the main analysis. Among
the three configurations considered here, it yields the largest number of
surviving expressions and corresponds to the range used in the reference
filtering procedure. This choice should not, however, be interpreted as
evidence that the physical filter is independent of the sampling range.

\section{Sensitivity of the fitting results to the ESR parameter bounds}\label{app:range-sensitivity}

The fitting stage requires finite bounds for the free ESR parameters.
To assess the dependence of the fitting results on this choice, we repeated
the parameter estimation using the bounds $[-5,5]$, $[-10,10]$, and
$[-20,20]$. In this test, the input sample was kept fixed: we used the same
786 candidates selected by the physical filter with trial parameter vectors
drawn from $[-5,5]$ and the DESI DR2+CMB $w_0w_a$CDM reference background.
Each candidate was then fitted independently using the three sets of ESR
parameter bounds. Only the bounds on the ESR parameters were varied; the
bounds on the cosmological parameters were kept unchanged, as described in
Section~\ref{sec:ParFit}.

The number of saved fit results decreases from 738 for $[-5,5]$ to 725
for $[-10,10]$ and 711 for $[-20,20]$. Relative to the 786 input candidates,
these values correspond to retained fractions of 93.89\%, 92.24\%, and
90.46\%, respectively. The fraction of saved fits with at least one ESR
parameter reaching a fitting bound also decreases as the bounds are widened,
from 13.0\% to 5.7\% and 2.1\% (Table~\ref{tab:bound-sensitivity}).

The minimum-MDL candidate is the same in all three runs: the expression
$f_{\rm ESR}(x)=\exp(\theta_0/x)$, corresponding to the normalized evolution
\begin{equation}
    X(z)=
    \exp\!\left[
        \theta_0\left(-1+\frac{1}{1+z}\right)
    \right],
\end{equation}
where $x=1+z$. The minimum MDL values are approximately 708.589 in all
three cases, with a maximum difference of approximately $3.3\times10^{-6}$
Thus, within this sample restricted to complexities 4 and 5, changing the fitting bounds
does not alter the identity of the minimum-MDL candidate and has a negligible
effect on its MDL value.

The constant-density expression in Complexity 1 is not included in this comparison.
Therefore, the minimum reported here is the minimum within the restricted
complexities 4 and 5 sample, not the minimum of the complete catalogue.

We retain $[-5,5]$ as the default fitting bounds because this configuration
has the highest retained fraction among the three tests and is computationally
less demanding than the wider alternatives. Nevertheless, this comparison
is restricted to complexities 4 and 5 and does not establish that the fitting results are
independent of the parameter bounds at higher symbolic complexities. 

\begin{table}[hbt]
\centering
\footnotesize
\caption{Sensitivity of the C4--C5 fitting results to the ESR parameter
bounds. All runs use the same 786 candidates selected by the physical filter
with trial parameter vectors drawn from $[-5,5]$. The retained fraction is
the number of saved fit results divided by the number of input candidates.
The boundary-hit fraction is the fraction of saved fits with at least one
ESR parameter at a fitting bound.}
\label{tab:bound-sensitivity}
\begin{tabular}{lrrr}
\toprule
Quantity & $[-5,5]$ & $[-10,10]$ & $[-20,20]$\\
\midrule
Input candidates & 786 & 786 & 786 \\
Saved fit results & 738 & 725 & 711\\
Retained fraction & 93.89\% & 92.24\% & 90.46\%\\
Boundary-hit fraction & 13.0\% & 5.7\% & 2.1\%\\
Minimum MDL & 708.589054847 & 708.589056568 & 708.589058151\\
Best tree index & 95 & 95 & 95\\
\bottomrule
\end{tabular}
\end{table}

\section{Fit-stage retention across symbolic complexities}
\label{app:fit-completion}

Table~\ref{tab:fit-completion} summarizes the number of expressions generated
by ESR, the number of candidates that passed the physical filter,
and the number of fit results retained after the
fitting stage, for each symbolic complexity from 1 to 8. 

The number of valid generated expressions increases rapidly with symbolic complexity, from 2 candidates at complexity 1 to 307,671 at complexity 8, resulting in 361,432 expressions across all the complexity levels considered. Of these, 179,331 passed the physical filter, corresponding to an overall survival fraction of   
49.62$\%$. 

Regarding the subsequent fitting stage, all candidates passing the physical filter 
are retained at complexities 1 to 3. At higher complexities, the retained fraction 
is generally lower, although the trend is not strictly monotonic: Complexity 5 has
a higher retained fraction than complexity 4.
The fraction decreases to 78.06$\%$ at complexity 8. This reduction reflects the
cumulative effect of the subsequent stages of the pipeline, which include
parameter estimation, physical validation, symbolic parameter reduction
and refitting, and final domain checks. 

Overall, 141,677 fit-result rows are retained from the 179,331
physical-filter survivors, corresponding to a fit-stage retention
fraction of 79.00$\%$. Relative to the initial set of generated
expressions, the final number of retained fit results corresponds
to 39.20$\%$ of the original candidate set. 

\begin{table}[t]
\centering
\footnotesize
\caption{ESR results per complexity for the DESI DR2+CMB
$w_0w_a$CDM reference-filter sample.}
\label{tab:fit-completion}
\begin{tabular}{lrrr}
\toprule
Complexity & ESR & Filter survivors & Saved fit-result rows\\
\midrule
C1 & 2 & 2 & 2 \\
C2 & 6 & 4 & 4 \\
C3 & 34 & 25 & 25 \\
C4 & 182 & 114 & 105 \\
C5 & 1,098 & 672 & 633\\
C6 & 6,930 & 3,954 & 3,401 \\
C7 & 45,509 & 24,086 & 20,052 \\
C8 & 307,671 & 150,474 & 117,455 \\
\midrule
Total & 361,432 & 179,331 & 141,677\\
\bottomrule
\end{tabular}
\end{table}

\bibliography{paper}

\begin{thebibliography}{}
\expandafter\ifx\csname natexlab\endcsname\relax\def\natexlab#1{#1}\fi
\providecommand{\url}[1]{\href{#1}{#1}}
\providecommand{\dodoi}[1]{doi:~\href{http://doi.org/#1}{\nolinkurl{#1}}}
\providecommand{\doeprint}[1]{\href{http://ascl.net/#1}{\nolinkurl{http://ascl.net/#1}}}
\providecommand{\doarXiv}[1]{\href{https://arxiv.org/abs/#1}{\nolinkurl{https://arxiv.org/abs/#1}}}

\bibitem[{Arjona \& Nesseris(2020)}]{Arjona_2020}
Arjona, R., \& Nesseris, S. 2020, Physical Review D, 101, 123525, \dodoi{10.1103/PhysRevD.101.123525}

\bibitem[{Barboza \& Alcaniz(2008)}]{Barboza2008}
Barboza, Jr., E.~M., \& Alcaniz, J.~S. 2008, Phys. Lett. B, 666, 415, \dodoi{10.1016/j.physletb.2008.08.012}

\bibitem[{Bartlett {et~al.}(2024)Bartlett, Desmond, \& Ferreira}]{Bartlett_ESR}
Bartlett, D.~J., Desmond, H., \& Ferreira, P.~G. 2024, IEEE Transactions on Evolutionary Computation, 28, 950.
\newblock \doarXiv{2211.11461}

\bibitem[{Berti {et~al.}(2025)Berti, Bellini, Bonvin, Kunz, Viel, \& Zumalacarregui}]{Berti2025}
Berti, M., Bellini, E., Bonvin, C., {et~al.} 2025, Physical Review D, 112, 023518, \dodoi{10.1103/dj3k-84v4}

\bibitem[{Chevallier \& Polarski(2001)}]{Chevallier2001}
Chevallier, M., \& Polarski, D. 2001, Int. J. Mod. Phys. D, 10, 213, \dodoi{10.1142/S0218271801000822}

\bibitem[{Cranmer(2023)}]{cranmer2023interpretable}
Cranmer, M. 2023, arXiv preprint arXiv:2305.01582

\bibitem[{de~Souza {et~al.}(2026)de~Souza, Sousa-Neto, Gonz{\'a}lez, \& Alcaniz}]{deSouza2026}
de~Souza, R., Sousa-Neto, A., Gonz{\'a}lez, J.~E., \& Alcaniz, J. 2026, Physical Review D, 113, 103531, \dodoi{10.1103/5zv5-yhwk}

\bibitem[{{DESI Collaboration} {et~al.}(2025){DESI Collaboration}, Abdul~Karim, {et~al.}}]{DESI2025}
{DESI Collaboration}, Abdul~Karim, M., {et~al.} 2025, Physical Review D, 112, 083515, \dodoi{10.1103/tr6y-kpc6}

\bibitem[{Dinda \& Maartens(2025)}]{Dinda2025}
Dinda, B.~R., \& Maartens, R. 2025, Journal of Cosmology and Astroparticle Physics, 2025, 120, \dodoi{10.1088/1475-7516/2025/01/120}

\bibitem[{Gr{\"u}nwald(2007)}]{Grunwald2007}
Gr{\"u}nwald, P.~D. 2007, The Minimum Description Length Principle, Adaptive Computation and Machine Learning (Cambridge, MA: MIT Press)

\bibitem[{Hoyt {et~al.}(2026)}]{Hoyt2026}
Hoyt, T.~J., {et~al.} 2026, Union3.1: Self-consistent measurements of host galaxy properties for 2000 Type Ia supernovae.
\newblock \doarXiv{2601.19424}

\bibitem[{Koza(1994)}]{Koza1994}
Koza, J.~R. 1994, Genetic Programming II: Automatic Discovery of Reusable Programs (Cambridge, MA: MIT Press)

\bibitem[{Lahav \& Shah(2026)}]{LahavShah2026}
Lahav, O., \& Shah, P. 2026, Nature Astronomy, 10, 1431, \dodoi{10.1038/s41550-026-02977-5}

\bibitem[{Linder(2003)}]{Linder2003}
Linder, E.~V. 2003, Physical Review Letters, 90, 091301, \dodoi{10.1103/PhysRevLett.90.091301}

\bibitem[{Makke \& Chawla(2024)}]{MakkeChawla2024}
Makke, N., \& Chawla, S. 2024, Artificial Intelligence Review, 57, \dodoi{10.1007/s10462-023-10622-0}

\bibitem[{Meurer {et~al.}(2017)Meurer, Smith, Paprocki, {\v{C}}ert{\'i}k, Kirpichev, Rocklin, Kumar, Ivanov, Moore, Singh, Rathnayake, Vig, Granger, Muller, Bonazzi, Gupta, Vats, Johansson, Pedregosa, Curry, Terrel, Rou{\v{c}}ka, Saboo, Fernando, Kulal, Cimrman, \& Scopatz}]{Meurer2017}
Meurer, A., Smith, C.~P., Paprocki, M., {et~al.} 2017, PeerJ Computer Science, 3, e103, \dodoi{10.7717/peerj-cs.103}

\bibitem[{{Planck Collaboration} {et~al.}(2020){Planck Collaboration}, Aghanim, Akrami, {et~al.}}]{Planck2018}
{Planck Collaboration}, Aghanim, N., Akrami, Y., {et~al.} 2020, Astronomy \& Astrophysics, 641, A6, \dodoi{10.1051/0004-6361/201833910}

\bibitem[{Popovic {et~al.}(2026)}]{Popovic2026}
Popovic, B., {et~al.} 2026, Monthly Notices of the Royal Astronomical Society, 548, stag632, \dodoi{10.1093/mnras/stag632}

\bibitem[{Schmidt \& Lipson(2009)}]{SchmidtLipson2009}
Schmidt, M., \& Lipson, H. 2009, Science, 324, 81, \dodoi{10.1126/science.1165893}

\bibitem[{Scolnic {et~al.}(2022)Scolnic, Brout, Carr, Riess, Davis, Dwomoh, Jones, Ali, Charvu, Chen, Peterson, Popovic, Rose, Wood, Brown, Chambers, Coulter, Dettman, Dimitriadis, Filippenko, Foley, Jha, Kilpatrick, Kirshner, Pan, Rest, Rojas-Bravo, Siebert, Stahl, \& Zheng}]{Scolnic2022}
Scolnic, D., Brout, D., Carr, A., {et~al.} 2022, The Astrophysical Journal, 938, 113, \dodoi{10.3847/1538-4357/ac8b7a}

\bibitem[{Sousa {et~al.}(2024)Sousa, Bartlett, Desmond, \& Ferreira}]{Sousa_2024}
Sousa, T., Bartlett, D.~J., Desmond, H., \& Ferreira, P.~G. 2024, Physical Review D, 109, \dodoi{10.1103/physrevd.109.083524}

\bibitem[{Sousa-Neto {et~al.}(2025)Sousa-Neto, Bengaly, Gonzalez, \& Alcaniz}]{Sousa-Neto:2025gpj}
Sousa-Neto, A., Bengaly, C., Gonzalez, J.~E., \& Alcaniz, J. 2025, Phys. Dark Univ., 50, 102108, \dodoi{10.1016/j.dark.2025.102108}

\bibitem[{Sousa-Neto {et~al.}(2026)Sousa-Neto, Landau, \& Alcaniz}]{Sousa-Neto:2026zsb}
Sousa-Neto, A., Landau, S.~J., \& Alcaniz, J. 2026.
\newblock \doarXiv{2610.27209}

\bibitem[{Udrescu \& Tegmark(2020)}]{UdrescuTegmark2020}
Udrescu, S.-M., \& Tegmark, M. 2020, Science Advances, 6, eaay2631, \dodoi{10.1126/sciadv.aay2631}

\bibitem[{Wang \& Freese(2006)}]{Wang2006}
Wang, Y., \& Freese, K. 2006, Physics Letters B, 632, 449–452, \dodoi{10.1016/j.physletb.2005.10.083}

\bibitem[{Wang \& Freese(2026)}]{WangFreese2026}
---. 2026, Journal of Cosmology and Astroparticle Physics, 2026, 023, \dodoi{10.1088/1475-7516/2026/02/023}

\bibitem[{Wang \& Garnavich(2001)}]{Wang2001}
Wang, Y., \& Garnavich, P.~M. 2001, The Astrophysical Journal, 552, 445, \dodoi{10.1086/320581}

\end{thebibliography}

\end{document}